\documentclass[aps,prl,reprint,groupedaddress,amsmath,amssymb]{revtex4-1}
\usepackage{graphicx}
\begin{document}

\title{Transfer matrix method for four-flux radiative transfer}

\author{Brian Slovick}\email{Corresponding author: brian.slovick@sri.com}
\author{Zachary Flom}
\author{Lucas Zipp}
\author{Srini Krishnamurthy}
\affiliation{Applied Optics Laboratory, SRI International, Menlo Park, California 94025, United States}
\begin{abstract}
We develop a transfer matrix formalism for four-flux radiative transfer models, which is ideally suited for studying transport through multiple scattering layers. The model, derived for spherical particles within the diffusion approximation, predicts the specular and diffuse reflection and transmission of multilayer composite films for diffuse or collimated incidence. The model shows remarkable agreement with numerical Monte Carlo simulations for a range of absorption and film thicknesses, as well as for an example multilayer slab.

\end{abstract}
\maketitle

\section{Introduction}
Modeling light propagation and scattering in multilayer media is relevant to a variety of applications, including atmospheric science \cite{Chandrasekhar2013,Ricchiazzi1998,Sokoletsky2014,Meador1980,Emde2016} and astronomy \cite{Whitney2003,Heng2014,Leenaarts2012}, remote sensing \cite{Platnick2000,Hedley2009,Kotchenova2006,Kotchenova2008}, and biomedical optics \cite{Tarvainen2005,Wang2007,Lubbers2014,Cheong1990}. For example, clouds and aerosols in the Earth's atmosphere \cite{Meador1980,Emde2016}, and the photosphere and chromosphere in the solar atmosphere \cite{Leenaarts2012}, can be modeled as multiple plane-parallel layers to study the impact of radiative transfer on climate and temperature variations. In remote sensing, plane-parallel radiative transfer models are used to correct for atmospheric scattering to improve image clarity \cite{Kotchenova2006,Kotchenova2008}, while in biomedical optics, layered models of blood \cite{Lubbers2014} and tissue \cite{Cheong1990} can be used to improve imaging techniques such as diffuse optical tomography used in neuroscience and cancer detection \cite{Edstrom2007,Eggebrecht2014}.

Several methods have been developed to solve the radiative transfer equation for the plane-parallel geometry. A common approach is to use Monte Carlo or raytracing simulations \cite{Wang2007,Iwabuchi2015}, in which the photon trajectory is modeled as a random walk with a step length and direction determined probabilistically by the mean free path and phase function, respectively. However, Monte Carlo methods are computationally expensive and time consuming, particularly for strongly scattering media. On the other hand, analytical models decompose the scattered light into a finite number of fluxes. The simplest model is the two-flux Kubelka-Munk model \cite{Kubelka1948,Vargas1997b}, which decomposes the flux into two isotropic diffuse fluxes propagating in the forward and backward directions perpendicular to the slab. The Kubelka-Munk model has found widespread use in paint design \cite{Orel1997}, radiation barriers \cite{Slovick2015}, and computer graphics \cite{Haase1992}. However, because it assumes diffuse incident light, it cannot predict specular reflectance and is only accurate for thick films \cite{Vargas1997b}, much thicker than the mean free path.

Alternatively, four-flux models decompose the total flux into collimated and diffuse components traveling in the forward and backward directions \cite{Ishimaru1997,Maheu1984,Vargas1998,Simonot2016}. In principle, four-flux models are more accurate for thin films in which there is a substantial unscattered component. Four-flux models have been applied recently to design thermal barrier coatings \cite{Wang2014}, x-ray shielding \cite{Haber2012}, dental composites \cite{Bayou2013}, and nanotechnology for energy applications \cite{Laaksonen2014}. However, a major drawback of the current formalism is that it cannot be easily extended to multiple layers \cite{Roze2001}. The reason for this can be traced to the way the interface reflections are included. In general, the  solution to the four-flux model contains four integration constants, which are determined by imposing boundary conditions on the flux values at the interfaces of the slab \cite{Ishimaru1997}. While this process is straightforward for a single layer, each additional layer adds four integration constants, and so the complexity increases rapidly with the number of layers.

In this article, we resolve this issue by developing a transfer matrix formalism for the four-flux model. We derive closed-form analytical expressions for the $4\times 4$ transfer matrices of both the film and interfaces. Multiplication of these matrices results in a total transfer matrix, from which the diffuse and specular reflection and transmission can be calculated. Our model is derived within the diffusion approximation, which assumes the diffuse flux is nearly isotropic \cite{Wang2007,Gemert1987}, and for spherical particles, though the formalism is applicable to other particle shapes. The model shows remarkable agreement with Monte Carlo simulations for a range of absorption values and film thicknesses, as well as for an example multilayer slab. The closed-form and accuracy of the model suggest it can be an efficient and reliable tool for studying radiative transfer for a variety of applications.

\section{Four-flux model}
In the four-flux model, the total flux in the film is decomposed into four fluxes: two collimated or specular fluxes traveling perpendicular to the slab in the forward and backward directions, $F_c^+(z)$ and $F_c^-(z)$, and two diffuse fluxes traveling in the forward and backward directions, $F_d^+(z)$ and $F_d^-(z)$ \cite{Ishimaru1997}. The collimated fluxes decay due to absorption and scattering as \cite{Ishimaru1997,Maheu1984,Vargas1998}
\begin{equation}
\frac{dF^+_c}{dz}=-(k+S_f+S_b)F^+_c,
\end{equation}
\begin{equation}
\frac{dF^-_c}{dz}=(k+S_f+S_b)F^-_c,
\end{equation}
where $k$ is the absorption coefficient and $S_f$ and $S_b$ are the scattering coefficients into the forward and backward hemispheres, respectively. The diffuse fluxes obey the following equations \cite{Ishimaru1997,Maheu1984,Vargas1998}:
\begin{equation}
\frac{dF^+_d}{dz}=-(K+S)F^+_d+SF_d^-+S_fF_c^++S_bF_c^-,
\end{equation}
\begin{equation}
\frac{dF^-_d}{dz}=(K+S)F^-_d-SF_d^+-S_fF_c^--S_bF_c^+,
\end{equation}
 where $K$ and $S$ are the absorption and scattering coefficients for the diffuse fluxes. The forward traveling diffuse flux decays due to absorption and scattering, and grows due to scattering of the backward traveling diffuse flux, forward scattering of the forward traveling collimated flux, and backscattering of the backward traveling collimated flux. The same process applies to the backward traveling diffuse flux. The solutions to Eqs. (1)-(4) are \cite{Ishimaru1997,Maheu1984}
\begin{equation}
F_c^+(z)=C_1 e^{-\lambda z}, \quad F_c^-(z)=C_4 e^{\lambda z}
\end{equation}
\begin{equation}
F_d^+(z)=C_1 A_1 e^{-\lambda z}+C_2e^{-\alpha z}+C_3 e^{\alpha z} +C_4 A_4 e^{\lambda z}
\end{equation}
\begin{equation}
F_d^-(z)=C_1 B_1 e^{\lambda z}+C_2 A_2 e^{-\alpha z}+C_3 A_3 e^{\alpha z} +C_4 B_4 e^{\lambda z}
\end{equation}
where $C_1$, $C_2$, $C_3$, and $C_4$ are integration constants determined by the boundary conditions and
$$
A_1=B_4=\frac{S S_b+S_f(K+S+\lambda)}{\alpha^2-\lambda^2},
$$
$$
A_2=A_3^{-1}=\frac{K+2S-\alpha}{K+2S+\alpha},
$$
$$
A_4=B_1=\frac{S S_f+S_b(K+S-\lambda)}{\alpha^2-\lambda^2},
$$
$$
\lambda=k+S_f+S_b, \quad \alpha=[k(K+2S)]^{1/2}.
$$
The forward and backward scattering coefficients for the collimated flux are related to the total scattering cross section $\sigma_s$ and the scattering cross section into the forward hemisphere $\sigma_f$ by \cite{Chylek1973,Vargas1997c}
$$
S_f=\sigma_f N, \quad S_b=(\sigma_s-\sigma_f)N,
$$
where $N$ is the particle density. Note that the total scattering coefficient is $S_f+S_b=\sigma_s N$. The absorption coefficient for the collimated flux is given by
$$
k=(\sigma_e-\sigma_s)N,
$$
where $\sigma_e$ is the extinction coefficient. The scattering and absorption coefficients for the diffuse flux $S$ and $K$ in general depend on the angular distribution of the flux. We assume the Eddington or diffusion approximation \cite{Wang2007,Gemert1987}, in which the diffuse flux is written as an isotropic term with an anisotropic cosine-dependent perturbation. In the diffusion approximation \cite{Gemert1987,Star1988}
$$
S=\frac{3}{4}(1-g)(S_f+S_b)-\frac{1}{4}k , \quad K=2k.
$$

\section{Transfer matrix}
First, we derive the transfer matrix for the film. To do so, it is necessary to write the flux values at the back interface ($z=d$) in terms of the flux values at the front interface ($z=0$). This procedure involves evaluating the fluxes in Eqs. (5)-(7) at $z=0$, solving for $C_1$, $C_2$, $C_3$, and $C_4$ and substituting these values into the fluxes evaluated at $z=d$. This results in the following transfer matrix equation:
\begin{equation}
\begin{bmatrix} 
F_c^+(d) \\
F_c^-(d) \\
F_d^+(d) \\
F_d^-(d) \\
\end{bmatrix}
=
\begin{bmatrix} 
M_{f,11}&M_{f,12}&M_{f,13}&M_{f,14}\\
M_{f,21}&M_{f,22}&M_{f,23}&M_{f,24}\\
M_{f,31}&M_{f,32}&M_{f,33}&M_{f,34}\\
M_{f,41}&M_{f,42}&M_{f,43}&M_{f,44}\\
\end{bmatrix}
\begin{bmatrix} 
F_c^+(0) \\
F_c^-(0) \\
F_d^+(0) \\
F_d^-(0) \\
\end{bmatrix},
\end{equation}
where the transfer matrix elements are given by
$$
M_{f,11}=e^{-\lambda d}, \quad M_{f,12}=0, \quad M_{f,13}=0, \quad M_{f,14}=0,
$$
$$
M_{f,21}=0, \quad M_{f,22}=e^{\lambda d}, \quad M_{f,23}=0, \quad M_{f,24}=0,
$$
$$
M_{f,31}=A_1 e^{-\lambda d}+\frac{A_1 (A_2 e^{\alpha d}-A_3 e^{-\alpha d})-2 B_1 \sinh(\alpha d)}{A_3-A_2},
$$
$$
M_{f,32}=A_4 e^{\lambda d}+\frac{A_4 (A_2 e^{\alpha d}-A_3 e^{-\alpha d})-2 B_4 \sinh(\alpha d)}{A_3-A_2},
$$
$$
M_{f,33}=\frac{A_3 e^{-\alpha d}-A_2 e^{\alpha d}}{A_3-A_2}, \quad M_{f,34}=\frac{2\sinh (\alpha d)}{A_3-A_2},
$$
$$
M_{f,41}=B_1 e^{-\lambda d}+\frac{2A_1 \sinh (\alpha d)+B_1(A_2 e^{-\alpha d}-A_3 e^{\alpha d})}{A_3-A_2},
$$
$$
M_{f,42}=B_4 e^{\lambda d}+\frac{A_3 e^{\alpha d}(A_2 A_4-B_4)-A_2 e^{-\alpha d}(A_3 A_4-B_4)}{A_3-A_2},
$$
$$
M_{f,43}=-\frac{2\sinh (\alpha d)}{A_3-A_2}, \quad M_{f,44}=\frac{A_3 e^{\alpha d}-A_2 e^{-\alpha d}}{A_3-A_2}.
$$

\begin{figure*}
\includegraphics[width=120mm]{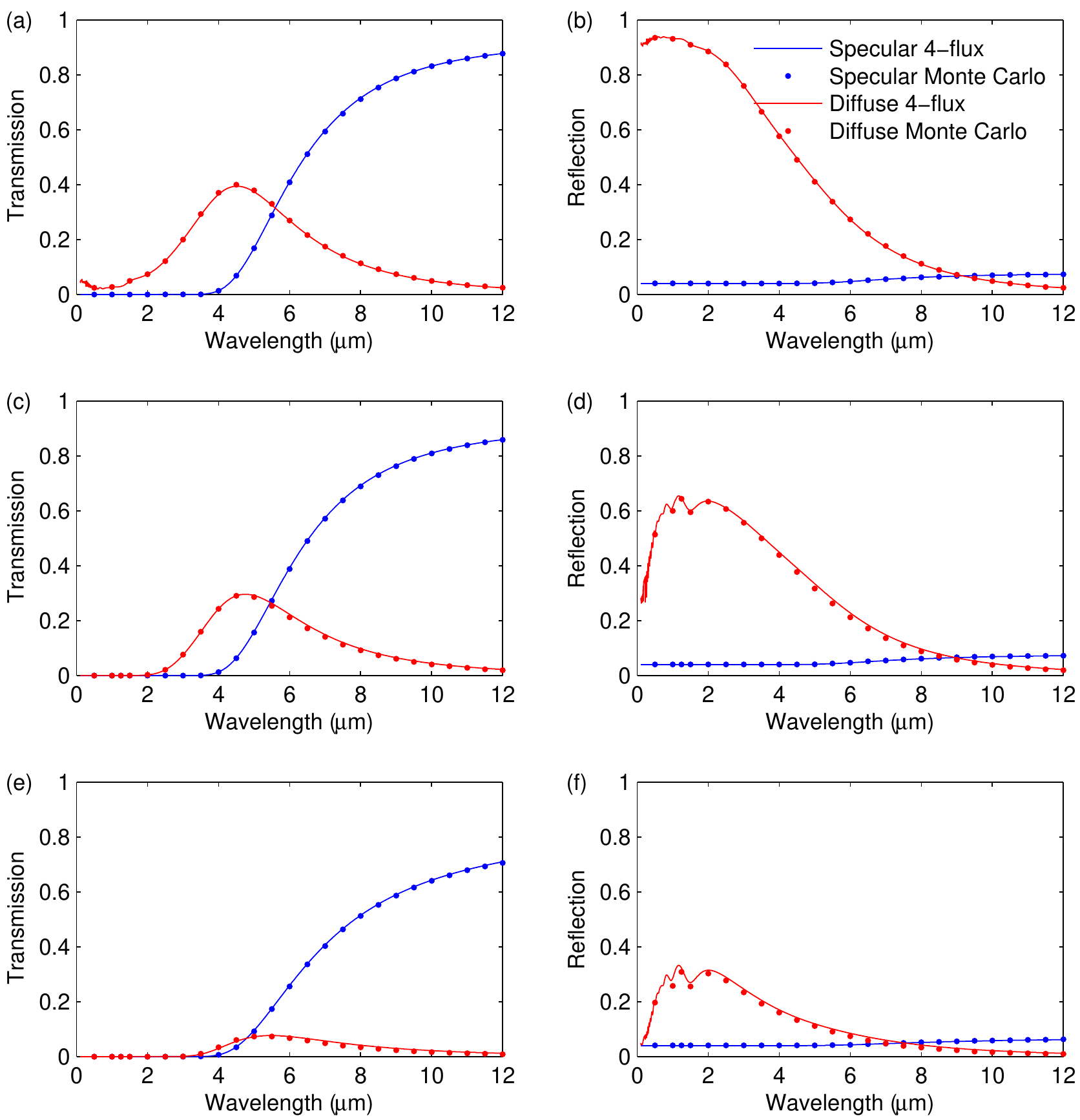}
\caption{\label{fig:epsart} Comparison of the specular and diffuse transmission and reflection calculated using the four-flux model (lines) and Monte Carlo simulations (dots) for a 100-$\mu$m thick slab with a matrix index of 1.5 and 30\% loading of 500-nm particles with a real refractive index of 2.5 and imaginary refractive index of zero [(a),(b)], $10^{-3}$ [(c),(d)], and $10^{-2}$ [(e),(f)].}
\end{figure*}

The transfer matrix for the entry and exit interfaces is obtained from the known values of the scattering matrix, which relates the outgoing fluxes to the incoming fluxes. Assuming the diffuse fluxes do not reflect or transmit into specular fluxes, and vice versa, the scattering matrix equation for the interfaces is
 $$
\begin{bmatrix} 
F_c^-(0) \\
F_c^+(d) \\
F_d^-(0) \\
F_d^+(d) \\
\end{bmatrix}
=
\begin{bmatrix} 
R_s&T_s&0&0\\
T_s&R_s&0&0\\
0&0&R_d^+&T_d^-\\
0&0&T_d^+&R_d^-\\
\end{bmatrix}
\begin{bmatrix} 
F_c^+(0) \\
F_c^-(d) \\
F_d^+(0) \\
F_d^-(d) \\
\end{bmatrix}
$$
where $T_s$ and $R_s$ are the specular Fresnel transmission and reflection coefficients for normal incidence and $T_d^{+-}$ and $R_d^{+-}$ are the transmission and reflection coefficients for the diffuse flux, where $+$ and $-$ denote incidence from $z=0$ and $z=d$, respectively. In the diffusion approximation, these coefficients are given by \cite{Haskell1994,Wang2007}
$$
R_d^{+-}=\frac{R_\phi^{+-}+R_J^{+-}}{2-R_\phi^{+-}+R_J^{+-}}, \quad T_d^{+-}=1-R_d^{+-},
$$
where
$$
R_\phi^{+-}=2 \int _0^{\pi/2} d\theta \sin(\theta) \cos(\theta) R^{+-}(\theta),
$$
$$
R_J^{+-}=3 \int _0^{\pi/2} d\theta \sin(\theta) \cos^2(\theta) R^{+-}(\theta),
$$
where $R^{+-}(\theta)$ is the polarization-averaged angular dependent Fresnel reflectivity \cite{Haskell1994,Wang2007}. Using the $S$ to $M$ matrix conversion procedure described in the appendix, the interface transfer matrix elements are
$$
M_{i,11}=\frac{T_s^2-R_s^2}{T_s}, \quad M_{i,12}=\frac{R_s}{T_s}, \quad M_{i,13}=M_{i,14}=0,
$$
$$
M_{i,21}=-\frac{R_s}{T_s}, \quad M_{i,22}=\frac{1}{T_s}, \quad M_{i,23}=M_{i,24}=0,
$$
$$
M_{i,31}=M_{i,32}=0, \quad M_{i,33}=\frac{T_d^+T_d^--R_d^+R_d^-}{T_d^-}, \quad M_{i,34}=\frac{R_d^-}{T_d^-},
$$
$$
M_{i,41}=M_{i,42}=0, \quad M_{i,43}=-\frac{R_d^+}{T_d^-}, \quad M_{i,44}=\frac{1}{T_d^-}.
$$
The primary advantage of the transfer matrix formalism is the ability to obtain an overall transfer matrix for a multilayer slab by multiplying the transfer matrices of the individual layers and interfaces $\textbf{M}_j$ as
$$
\textbf{M}=\prod_{j=N}^1 \textbf{M}_j.
$$
For example, the overall transfer matrix of a single film with two similar interfaces is given by $\textbf{M}_i \textbf{M}_f \textbf{M}_i$. The diffuse and specular reflection and transmission are then calculated using the $M$ to $S$ matrix conversion procedure described in the appendix.

\begin{figure*}
\includegraphics[width=120mm]{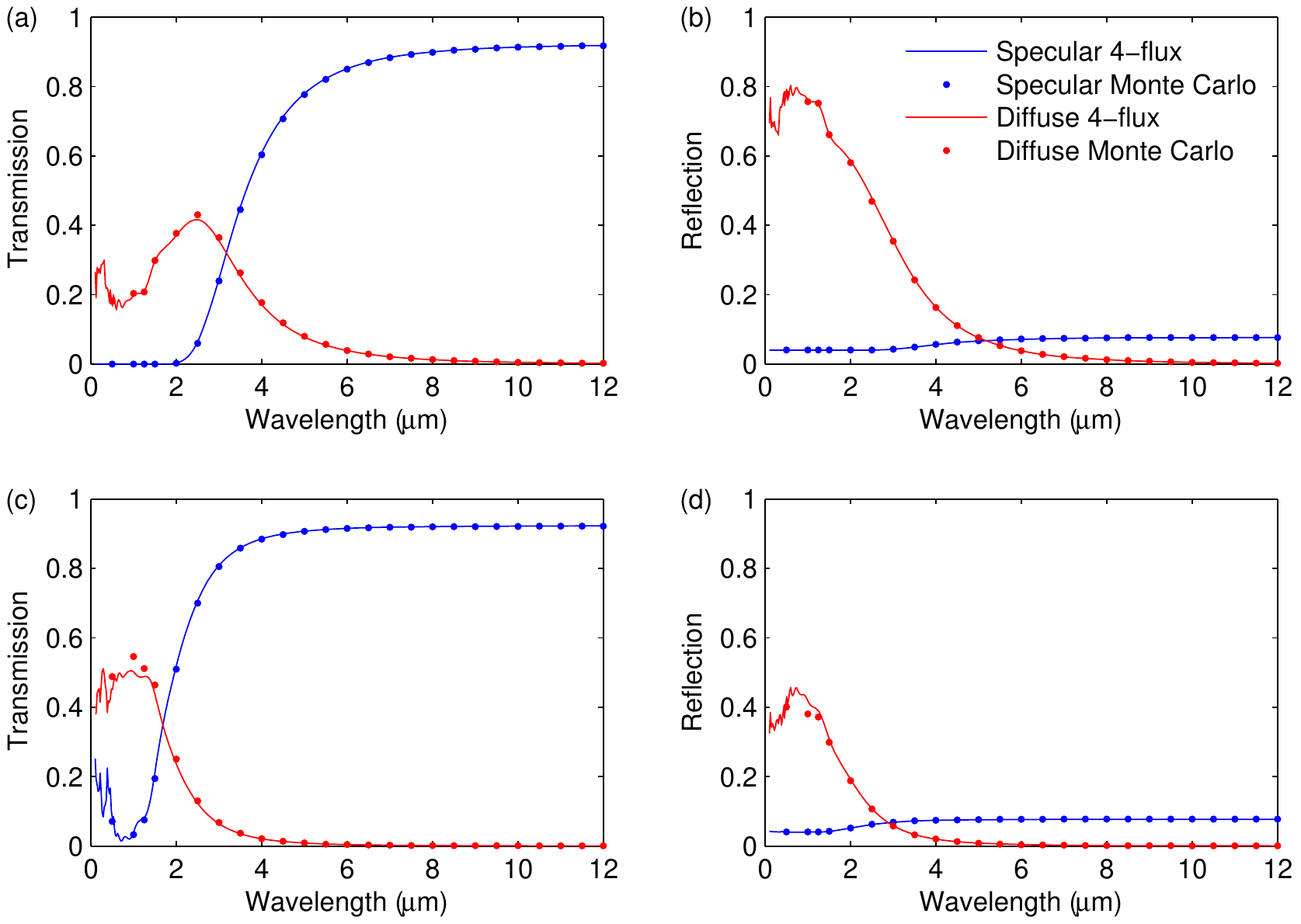}
\caption{\label{fig:epsart} Comparison of the specular and diffuse transmission and reflection calculated using the four-flux model (lines) and Monte Carlo simulations (dots) for a slab with a matrix index of 1.5 and 30\% loading of 500-nm particles with a real refractive index of 2.5 and slab thicknesses of 10 $\mu$m [(a),(b)] and 1 $\mu$m [(c),(d)].}
\end{figure*}

\section{Monte Carlo Validation}
To validate the four-flux transfer matrix model, we performed Monte Carlo simulations using the commercial raytracing software LightTools by Optical Research Associates. All simulations trace 60,000 rays with a relative ray power threshold of $10^{-5}$ and employ a probabilistic ray split at the interfaces. The first case we considered is intended to evaluate the accuracy of the four-flux model for different amounts of absorption in the particles. Figure 1 shows the calculated specular and diffuse transmission and reflection for a 100-$\mu$m thick slab with a matrix index of 1.5 and 30\% loading of 500-nm particles with a real refractive index of 2.5 and different values of the imaginary refractive index.

In the case of zero absorption, shown in Fig. 1(a) and (b), there is remarkable agreement between the models. As the absorption increases, shown in Fig. 1 (c)-(f), the agreement remains excellent but decreases slightly, the most notable difference being the slightly larger diffuse reflection predicted by the four-flux model. The likely reason for this can be traced to the assumed angular distribution of the diffuse flux in the four-flux model. The four-flux model assumes the diffuse flux is nearly isotropic, and thus generally more diffuse than the actual flux. As a result, the average penetration depth is shallower in the four-flux model, leading to less absorption and more diffuse reflection.

Next, we consider the accuracy of the model for different film thicknesses, assuming no absorption. In general, we expect the four-flux model to be more accurate for thicker films, as the assumption of nearly isotropic diffuse flux is valid. Figure 2 shows the calculated specular and diffuse transmission and reflection for the composite in Fig. 1 for slab thicknesses of 10 $\mu$m [(a),(b)] and 1 $\mu$m [(c),(d)]. Overall, we find excellent agreement between the four-flux model and the Monte Carlo simulations, though, as expected, the accuracy of the four-flux model decreases as the film thickness decreases. In particular, in the strong Mie scattering region around 1 $\mu$m the four-flux model predicts higher diffuse reflection and lower diffuse transmission. This can be explained by the directionality of the scattering in the Mie region. When the scattering is directional and the film is thin compared to the mean free path, the scattered flux will be nearly collimated, which contradicts the assumption of nearly isotropic diffuse flux in the four-flux model.

Finally, to illustrate the functionality of the transfer matrix formalism, we considered a two-layer composite consisting of two 50-$\mu$m slabs with 30\% loading of particles with an index of 2.5. The matrix indices and particle sizes in the first and second layers are 1.5 and 2 and 500 nm and 1000 nm, respectively. Figure 3 shows a comparison of the four-flux model to the Monte Carlo simulation for the multilayer film. Again we find excellent agreement between the two models, though the four-flux model predicts slightly higher diffusion reflection and lower diffuse transmission for wavelengths between 4 and 8 $\mu$m, which can again be explained by the assumption of a nearly isotropic diffuse flux in the four-flux model.  

\begin{figure}
\includegraphics[width=60mm]{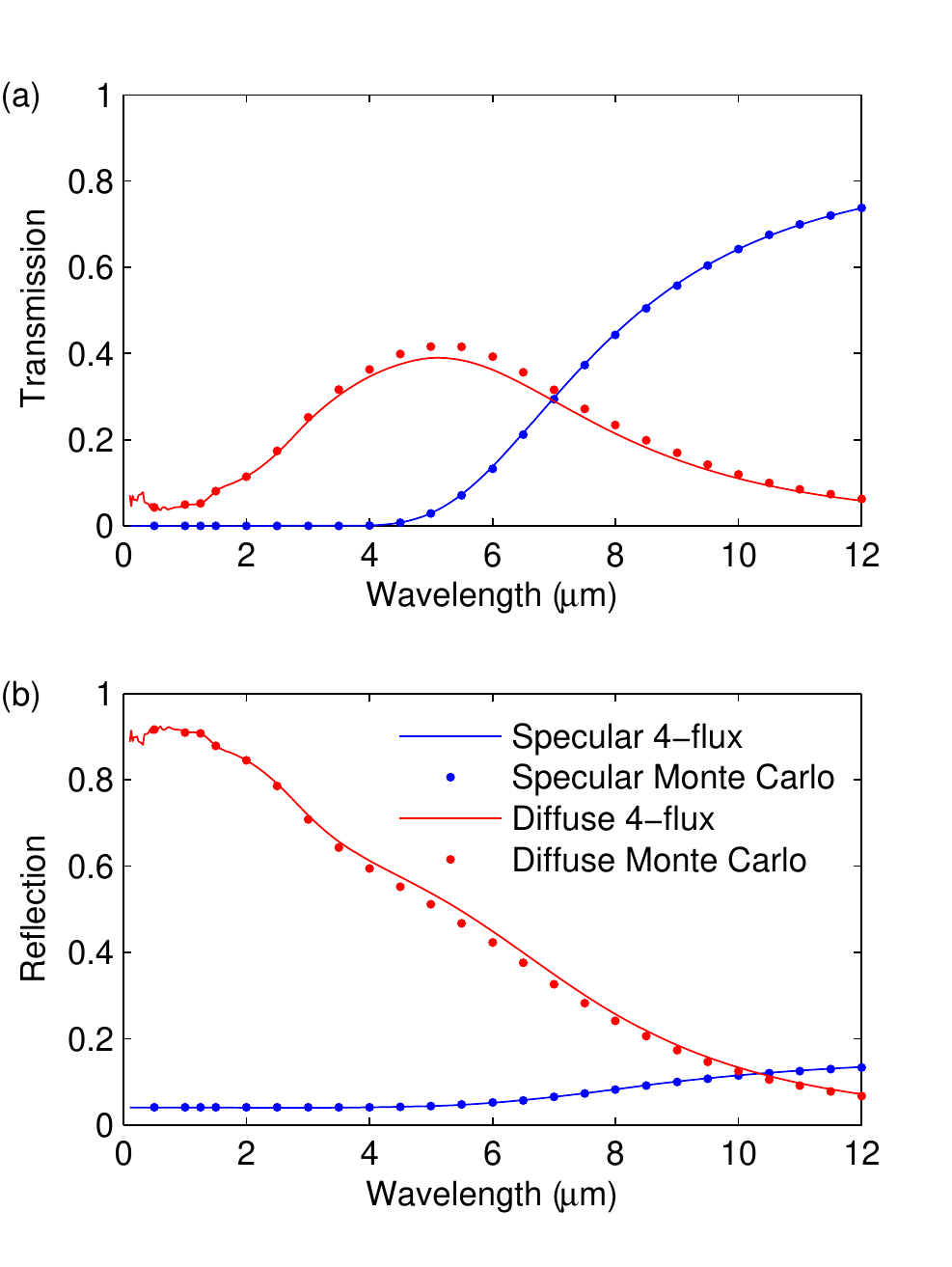}
\caption{\label{fig:epsart} Comparison of the specular and diffuse transmission and reflection calculated using the four-flux model (lines) and Monte Carlo simulations (dots) for a two-layer slab. Both layers are 50-$\mu$m thick and have 30\% loading of particles with a refractive index of 2.5. The matrix indices and particle sizes in the first and second layers are 1.5 and 2 and 500 nm and 1000 nm, respectively.}
\end{figure}

\section{Summary}
We have developed a closed-form transfer matrix formalism for the four-flux theory of radiative transfer, which is ideally suited for studying multilayered turbid media. The model provides expressions for both the transfer matrix of the inhomogeneous film as well as the interfaces. The model has been derived within the diffusion limit, which approximates the scattered flux as nearly isotropic, and for spherical particles, though the formalism is generally applicable to other shapes provided that the scattering parameters can be calculated. The model shows remarkable agreement with numerical Monte Carlo simulations for a wide range of absorption values and film thicknesses, as well as for an example multilayer slab. The closed form and accuracy of the model suggest it can be an efficient and reliable tool for studying radiative transfer for a variety of applications.

\section{acknowledgment}
This work was funded by the Materials and Manufacturing Directorate of the Air Force Research Laboratory (prime contract \# FA8650-11-D-5400) through UES subcontract \#S-925-004-0013.

\section{Appendix}
\subsection{Conversion between S and M matrix}
In this section, we provide the procedure to convert from the $M$ matrix to the $S$ matrix, whose elements correspond to the reflection and transmission coefficients. By definition, the $S$ matrix relates the outgoing fluxes to the incoming fluxes as
\begin{equation}
\begin{bmatrix} 
F_c^-(0) \\
F_c^+(d) \\
F_d^-(0) \\
F_d^+(d) \\
\end{bmatrix}
=
\begin{bmatrix} 
S_{11}&S_{12}&S_{13}&S_{14}\\
S_{21}&S_{22}&S_{23}&S_{24}\\
S_{31}&S_{32}&S_{33}&S_{34}\\
S_{41}&S_{42}&S_{43}&S_{44}\\
\end{bmatrix}
\begin{bmatrix} 
F_c^+(0) \\
F_c^-(d) \\
F_d^+(0) \\
F_d^-(d) \\
\end{bmatrix}.
\end{equation}
To obtain the $S$ matrix elements in terms of the $M$ matrix elements, Eq. (8) is solved for the outgoing fluxes as a function of the incoming fluxes. The $S$ matrix elements are then obtained by equating the coefficients of the incoming fluxes those in Eq. (9). We demonstrate this procedure by calculating the collimated and diffuse transmission and reflection coefficients for collimated incidence. These expressions are considerably simplified when calculated for light incident from the $z=d$ side of the slab, in which case
$$
T_{cc}=S_{12}=\frac{M_{44}}{M_{22} M_{44}-M_{24} M_{42}},
$$
$$
R_{cc}=S_{22}=\frac{M_{12}M_{44}}{M_{22} M_{44}-M_{24} M_{42}},
$$
$$
T_{cd}=S_{32}=\frac{-M_{42}}{M_{22} M_{44}-M_{24} M_{42}},
$$
$$
R_{cd}=S_{42}=\frac{M_{32} M_{44}-M_{34} M_{42}}{M_{22} M_{44}-T_{24} M_{42}},
$$
where, for example, $T_{cd}$ corresponds to diffuse transmission for collimated incidence. A similar procedure can be followed to convert the $S$ matrix elements to $M$ matrix elements. In this case, Eq. (9) is solved to obtain expressions for the fluxes at $z=d$ as a function of the $S$ parameters, which can be compared to Eq. (8) to determine the $M$ matrix elements. 
\subsection{Mie Scattering parameters}
Mie theory provides an exact solution for the scattering parameters of spherical particles in terms of the multipole Mie scattering coefficients \cite{Bohren2008}
$$
a_n=\frac{n_p\psi'_n(k_m a)\psi_n(k_p a)-n_m \psi_n(k_m a)\psi'_n(k_p a)}{n_p \xi'_n(k_m a)\psi_n(k_p a)-n_m\xi_n(k_m a)\psi'_n(k_p a)},
$$
$$
b_n=\frac{n_m \psi'_n(k_m a)\psi_n(k_p a)-n_p \psi_n(k_m a)\psi'_n(k_p a)}{n_m \xi'_n(k_m a)\psi_n(k_p a)-n_p \xi_n(k_m a)\psi'_n(k_p a)},
$$
where $n_p$ and $n_m$, respectively, are the refractive index of the particle and matrix, $k_{m,p}=n_{m,p} 2\pi/\lambda$, $\psi_n(x)=xj_n(x)$ and $\chi_n(x)=-xy_n(x)$ are the Riccati-Bessel functions where $j_n(x)$ and $y_n(x)$ are the spherical Bessel functions, and $\xi_n(x)=\psi_n(x)-i\chi_n(x)$. The primes denote differentiation with respect to the argument. The scattering cross section is given by
\begin{equation}
\sigma_s=\frac{2\pi}{k_1^2} \sum_{n=1}^\infty (2n+1) (|a_n|^2+|b_n|^2),
\end{equation}
and the extinction cross section by
\begin{equation}
\sigma_e=\frac{2\pi}{k_1^2} \sum_{n=1}^\infty (2n+1) \text{Re}(a_n+b_n).
\end{equation}
The anisotropy parameter, which measures the average cosine of the scattering angle, is given by \cite{Wang2007}
\begin{eqnarray}
g=\frac{4\pi}{k_1^2}\sum_{n=1}^\infty \frac{n(n+2)}{n+1}&& \left[ \frac{}{}\text{Re}(a_n a^*_{n+1}+b_n b^*_{n+1}) \right. \nonumber \\
&& \left. +\frac{2n+1}{n(n+1)}\text{Re}(a_nb^*_n) \right].
\end{eqnarray}
Finally, the scattering cross section for light into the forward hemisphere is given by \cite{Chylek1973,Vargas1997c}
\begin{eqnarray}
\sigma_f=\frac{1}{2}\sigma_s&&-\frac{2\pi}{k_1^2} \sum_{m''=2}^{\infty} \sum_{n'=1}^\infty p_{nm} \text{Re}(a_ma^*_n+b_mb^*_n) \nonumber \\
&&-\frac{2\pi}{k_1^2} \sum_{m'=1}^\infty \sum_{n'=1}^\infty q_{mn} \text{Re}(a_m b^*_n),
\end{eqnarray}
where the prime and double prime, respectively, denote odd and even integers, and
$$
p_{nm}=(-1)^{(m+n-1)/2}\frac{(2m+1)(2n+1)(m-1)!!n!!}{(m-n)(m+n+1)m!!(n-1)!!},
$$
$$
q_{nm}=(-1)^{(m+n)/2}\frac{(2m+1)(2n+1)m!!n!!}{m(m+1)n(n+1)(m-1)!!(n-1)!!}.
$$

\bibliography{bib}

\end{document}